# Sensitivity of topological edge states in a non-Hermitian dimer chain

Zhiwei Guo[*], Tengzhou Zhang, Juan Song, Haitao Jiang, and Hong Chen[*]

*Key Laboratory of Advanced Micro-structure Materials, MOE, School of Physics Science and Engineering, Tongji University,*

*Shanghai 200092, China*

## Abstract

Photonic topological edge states in one-dimensional dimer chains have long been thought to be robust to structural perturbations by mapping the topological Su-Schrieffer-Heeger model of a solid-state system. However, the edge states at the two ends of a finite topological dimer chain will interact as a result of near-field coupling. This leads to deviation from topological protection by the chiral symmetry from the exact zero energy, weakening the robustness of the topological edge state. With the aid of non-Hermitian physics, the splitting frequencies of edge states can be degenerated again and topological protection recovered by altering the gain or loss strength of the structure. This point of coalescence is known as the exceptional point (EP). The intriguing physical properties of EPs in topological structures give rise to many fascinating and counterintuitive phenomena. In this work, based on a finite non-Hermitian dimer chain composed of ultra-subwavelength resonators, we propose theoretically and verify experimentally that the sensitivity of topological edge states is greatly affected when the system passes through the EP. Using the EP of a non-Hermitian dimer chain, we realize a new sensor that is sensitive to perturbation at the end of the structure and yet topologically protected from internal perturbation. Our demonstration of a non-Hermitian topological structure with an EP paves the way for the development of novel sensors that are not sensitive to internal manufacturing errors but are highly sensitive to changes in the external environment.

[*] Corresponding author: Email: 2014guozhiwei@tongji.edu.cn, hongchen@tongji.edu.cn

## I. INTRODUCTION

Topological insulators, a hot research topic in physics, have greatly improved understanding of the classification of states in condensed matter physics. The fully occupied electronic band structure has the topological characteristics identified by the topological invariants [1]. Topological insulators have also opened up a new research stream in the development of new semiconductor devices to be used in quantum computing, high-fidelity quantum communication, and so on [2, 3]. Inspired by the topological properties of electronic band structures, scientists designed a photonic counterpart and observed the charming photonic edge states in the artificial photonic structures [4–6]. It is of great scientific significance to use topology to control the motion of photons, and this unique research has been extended to quasiperiodic systems [7-9]. Photonic topological edge states can overcome the scattering losses caused by structural defects and disorders and realize topologically protected photonic devices, such as unidirectional waveguides and single-mode lasers [4–6].

As one of the simplest topological structures, the one-dimensional (1D) dimer chain has been widely used in the study of photonic topological excitation. In this structure, the topological invariant can be directly identified by comparing the relative magnitude of intra-cell and inter-cell coupling coefficients [10]. Specifically, the edge states will appear symmetrically at two ends of the finite topological chain [11], and the topological order—winding number—will be directly observable in the microwave regime [12]. At present, research on the advantages of 1D dimer chains has been extended to nonlinear [13–17] and active [18–21] systems. However, non-Hermitian topological photonics is also a very hot research topic in topological physics. Advances in the field of non-Hermitian photonics based on parity-time (PT) symmetry have greatly improved the ability to design new photonic topological insulators in previously inaccessible ways [22, 23]. In general, the eigenvalues of open optical non-Hermitian systems are generally complex. However, the PT symmetric structure with real eigenvalues belongs to a very special non-Hermitian system. The most noticeable feature of a non-Hermitian system is that there are degenerate points in the Riemannian surface of the parameter space, where eigenvalues and corresponding eigenvectors simultaneously coalesce [24, 25]. These non-Hermitian degeneracies, also called exceptional points (EPs), have turned out to be the origin of many counterintuitive phenomena, such as loss-induced transparency [26], band merging [27], dynamic wireless energy transmission [28, 29], the chirality-reversal phenomenon [30], and mode transfer [31–34]. EPs provide a new way to design new highly sensitive sensors beyond the linear response, and performance can be optimized as the order of the EPs increases [35–45]. Non-Hermitian topological structures contain new physical phenomena not usually found in normal Hermitian topological structures [46–56]. To date, the unique topological order [48, 51], phase transition [46, 53, 54, 56], and edge state [47, 48, 50, 52] of the non-Hermitian 1D dimer chain have been studied. For example, the non-Bloch winding number as the topological invariant has been theoretically proposed [49], and topological protection of the edge states has been demonstrated experimentally [48]. Topological non-Hermitian systems provide an effective avenue for studying the intriguing properties of topological photonics involving EPs and developing new functional devices.

Near-field mode coupling is a fundamental physical effect that plays an important role in controlling electromagnetic waves [57, 58]. Researchers who have studied the near-field coupling of topological edge states have found many interesting phenomena, such as robust topological Fano resonance [59, 60] and Rabi splitting [61]. Specifically, in a finite non-Hermitian dimer waveguide array, the coupling effect of edge states leads to deviation from the topological zero mode and thus weakens the robustness of the edge states [62, 63]. In general, mode splitting induced by near-field coupling can be eliminated by increasing the length of the chain [57]. To recover topological protection, the coupling of the two edge states must be significantly reduced by increasing the chain length, which will cause the splitting edge state modes to return to zero energy. However, in a non-Hermitian system, the splitting frequencies can be degenerated again at the EPs by directly altering the gain or loss strength while keeping the length of the chain unchanged [62, 63]. Although the effects of near-field coupling on the robustness of topological edge states have been confirmed qualitatively from field distributions [62, 63], the behavior of novel EPs in non-Hermitian topological systems has not been reported. Topological edge states are generally thought to be robust to structural perturbations, as they result from nonlocal response based on the bulk-boundary correspondence. In contrast, the EP is often used to achieve highly sensitive sensors and is sensitive to slight variation in the environment. Thus, a question naturally arises: Can topological edge states be used to design new highly sensitive sensors by combining EPs?

In this work, we study experimentally the properties of the EP in a finite non-Hermitian topological dimer chain. The coupling between two edge states is presented, which is particularly relevant to the realization of EPs. By adding loss and gain to both ends of the dimer chain, we can obtain the non-Hermitian topological chain that satisfies PT symmetry and then observe the EP by increasing the loss or gain of the system. Moreover, we also study the sensitivity of topological edge states to disturbances in the environment before and after the EP. As a result, a new highly sensitive sensor with topological protection is realized based on the EP of topological edge states. In sharp contrast to traditional sensors, this new sensor based on non-Hermitian and topological characteristics has unique advantages. It is immune from internal disturbances of the structure and very sensitive to changes in the external environment at both ends. By combining non-Hermitian systems with topology photonics, we design a sensor that has both the robustness of topology and the sensitivity of EPs. In addition, an even more sensitive topological sensor could be designed in the future considering the higher order EPs realized by the synthetic dimension [64, 65]. Our findings not only present a novel photonic sensor with topological protection but also may be very useful to a variety of applications with non-Hermitian properties, including wireless energy transmission, switching, and filters.

**II. THE EPs OF EDGE STATES IN A FINITE NON-HERMITIAN DIMER CHAIN**

We consider a finite non-Hermitian topological dimer chain consisting of subwavelength resonators. The full circuit model of the composite resonator is shown in Fig. 1(a). The resonator is composed of a fundamental LC resonator, a negative resistance convertor (NIC) component, and a tunable resistor. Specifically, the LC resonator is constructed with a double-side winding structure and top and bottom layers connected by metal vias, as shown in Fig. 1(b). The composite resonator is fabricated on a commercial printed-circuit substrate, FR-4 ($\varepsilon_r$ = 4.75, $\tan\delta$ = 0.03), with a thickness of $h$ = 1.6 mm. The width and gap of the metal lines are $w$ = 1.12 mm and $g$ = 0.39 mm, respectively. The resonance frequency of the LC resonator is $\omega_0$ = 6.15 MHz, with $L_0$ = 1.12 μH, $C_0$ = 600 pF. The resonance frequency can be flexibly tuned by adding different lumped capacitors into the welding position, which is marked by the green rectangles in Fig. 1(b). At the resonance frequency, the electromagnetic field is mostly confined within the resonators; thus, the near-field coupling of resonators can be treated as the tight-binding model. In this work, modulation of gain and loss in the composite resonator is realized by the NIC component and tunable resistor, respectively. A metal-oxide-semiconductor field-effect transistor (MOSFET) is used to provide the effective gain. The circuit model and the connection schematic of the NIC component are shown in Fig. 1(c) and 1(d), respectively. The source of two MOSFETs is in the same ground, whereas their gates are connected to the drain of the opposite transistor. This is equivalent to a voltage-controlled current source whose input and output ports are in inverse parallel, thus playing a role in negative resistance. RF chocks ($L_R = 15$ μH) play a role in constant current while avoiding the influence of the alternating current (AC) signal on the direct current (DC) stabilized voltage source [66].

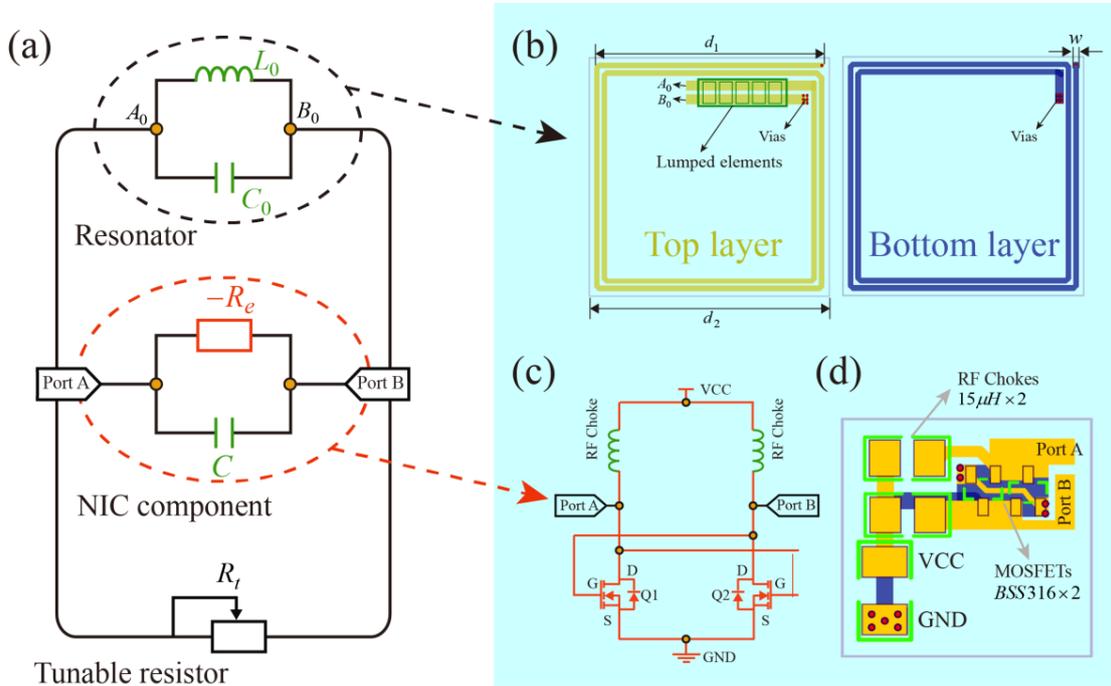

FIG. 1. **Composite resonator with tunable gain and loss designed in the current study.** (a) Effective circuit model of the composite resonator, composed of a simple LC resonator, a negative resistance convertor (NIC) component, and a tunable resistor. (b) Details of the composite resonator, where the gold and blue structures indicate the top and bottom copper layers, respectively. Here $d_1 = 46.2$ mm, $d_2 = 48$ mm, and $w = 1.12$ mm; the thickness of the substrate is $h = 1.6$ mm. The lumped circuit elements and vias are marked by the green rectangles and red dots, respectively. (c) Circuit model of the NIC component. The effective gain is tuned by the external direct current (DC) voltage source. (d) Schematic of the realization of the NIC component based on RF chokes and Metal-oxide-semiconductor field-effect transistors (MOSFETs).

The tunable gain and loss in the composite resonator can be realized by changing the external bias voltage on the NIC component and the resistance of the tunable resistor, respectively. First we decrease the resistance of the tunable resistor gradually without applying the external bias voltage. From the measured reflection spectrum in Fig. 2(a), it can be seen that with the decrease in resistance, the minimum value of the reflection spectrum increases gradually, and the half height and width of the spectrum become wider. This means that the effective loss is introduced into the resonator when the resistor is added to the circuit. The loss of the system is inversely proportional to the value of the shunted resistor. Next the external bias voltage is applied to the NIC component while the resistance is fixed at $R = 2.8$ kΩ. The corresponding reflection spectrum of the composite resistor is shown in Fig. 2(b). Owing to the effects of parasitic capacitance in the NIC component, the resonance frequency of the composite resonator shifts slightly. Overall, as the bias voltage increases, the minimum value of the reflection spectrum decreases gradually, and the half height and width of the spectrum become narrower, as shown in Fig. 2(b). Therefore, the effective gain is introduced into the resonator when the bias voltage is applied to the circuit. From Fig. 2, we can see that both the loss and the gain can be flexibly controlled in the composite circuit–based resonator.

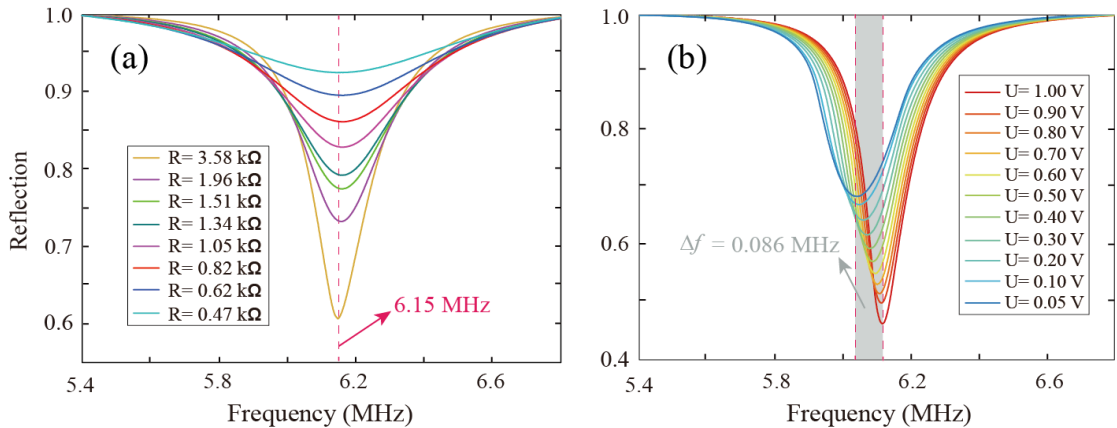

FIG. 2. **Measured reflection spectrum of the composite resonator.** (a) The reflection spectrum from changing the resistance without the external bias voltage. The resonant frequency, which is marked by the pink dashed line, is almost unchanged. (b) Similar to (a), but the external voltage changes while the resistance is fixed at $R = 2.8$ kΩ. The slight frequency shift of 0.086 MHz is marked by gray shading.

The 1D non-Hermitian topological dimer chain is shown schematically in Fig. 3(a). This chain involves six resonators: two on either end with gain and loss separated by four inner neutral resonators. To illustrate this more clearly, we show the amplifying, neutral, and lossy resonators in pink, green, and blue, respectively. The topological order of the 1D dimer chain is related to the relative magnitude of intra-cell ($\kappa_1$) and inter-cell ($\kappa_2$) coupling coefficients [11, 12]. Specifically, the Zak phases of the topological structure ($\kappa_1 < \kappa_2$) and trivial structure ($\kappa_1 > \kappa_2$) are $\pi$ and 0, respectively. Because of the bulk-boundary correspondence, the edge states will appear symmetrically at two ends of the chain for the finite topological structure. However, they will interact each other as a result of the near-field coupling. This will lead to deviation from the topological zero mode protected by the chiral symmetry from the exact zero energy, weakening the robustness of the topological edge state. The magnitude of intra-cell and inter-cell coupling coefficients are $\kappa_1$ = -0.4 MHz and $\kappa_2$ = -0.52 MHz, respectively. Based on the tight binding model, the associated $6 \times 6$ non-Hermitian Hamiltonian can be written as follows:

$$H = \begin{pmatrix} \omega_G & \kappa_1 & 0 & 0 & 0 & 0 \\ \kappa_1 & \omega_N & \kappa_2 & 0 & 0 & 0 \\ 0 & \kappa_2 & \omega_N & \kappa_1 & 0 & 0 \\ 0 & 0 & \kappa_1 & \omega_N & \kappa_2 & 0 \\ 0 & 0 & 0 & \kappa_2 & \omega_N & \kappa_1 \\ 0 & 0 & 0 & 0 & \kappa_1 & \omega_L \end{pmatrix}, \quad (1)$$

where $\omega_G = \omega_0 + ig_G - i\Gamma$, $\omega_N = \omega_0 - i\Gamma$, and $\omega_L = \omega_0 - ig_L - i\Gamma$. $g_G$ and $g_L$ denote the effective gain and loss of the resonators, respectively. Without considering the loss term $\Gamma = 0.03$ MHz in the neutral resonator and $g_G = g_L$, the non-Hermitian chain described by Eq. (1) can be seen as an ideal PT system. When a small frequency perturbation $\varepsilon$ affects the lossy resonator on the right of the dimer chain, the term $\omega_L$ becomes $\omega_L = \omega_0 + \varepsilon - ig_L - i\Gamma$. Without the loss of generality, we first establish an equivalent ideal PT system with balanced loss and gain on two ends of the chain, then obtain the EP by increasing the loss. To avoid the influence of vibration from the NIC component on the experiment when the external bias voltage is large, we select a suitable voltage on the gain resonator of $U = 1.3$ V. When the shunted resistor of the lossy resonator is $R = 10$ k$\Omega$, we can establish an ideal PT system with $g_G = g_L = 0.18$ MHz. Then we retain $g_G = 0.18$ MHz and increase $g_L$ to realize the EP. Using Eq. (1), we calculate the real eigenfrequencies dependent on parameter $g_L$ and frequency detuning $\varepsilon$, as shown in Fig. 3(b). It is important to note that the topological edge states can be identified in the band gap. The EP in this non-Hermitian system without perturbations is marked by the black circle. To see clearly the evolution of eigenfrequencies on parameter $g_L$, we show the enlarged eigenfrequencies of two edge states as a function of parameter $g_L$ without frequency detuning ($\varepsilon = 0$) in Fig. 3(c). As $g_L$ increases, the two splitting edge states gradually coalesce in the EP. We take the case of $g_L = 0.18$ MHz (marked by the red dashed line) as an example; the two edge states are represented by $\omega_+$ and $\omega_-$. The normalized wave functions of the two edge states $\omega_+$ and $\omega_-$ are shown in Fig. 3(d) and 3(e), respectively. From Fig. 3, we can see that the near-field coupling between the two edge states is a global response. Specifically, asymmetric and symmetric distributions are realized by the edge states $\omega_+$ and $\omega_-$, which means the coupling sign between the two edge states is positive, although the coupling sign between different resonators is negative.

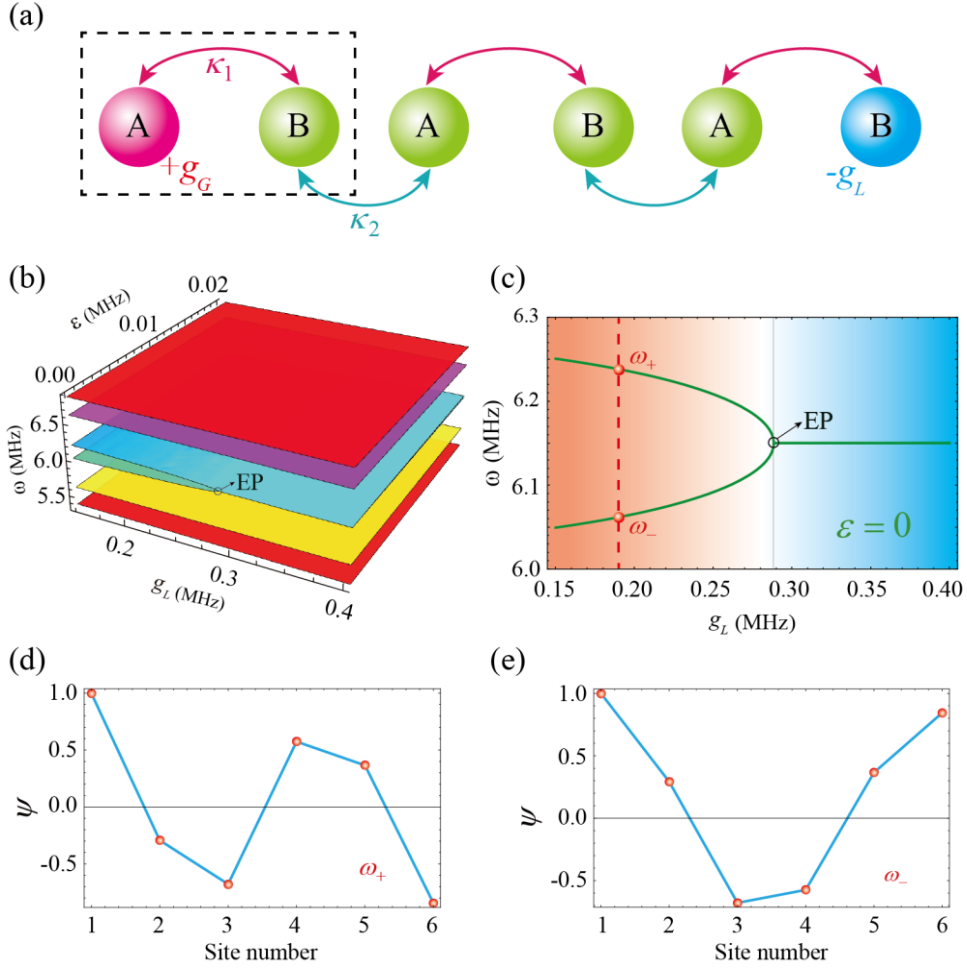

FIG. 3. **1D non-Hermitian topological dimer chain.** (a) Schematic of a topological dimer chain with six resonators. Effective loss and gain are added into the left and right resonators, respectively. (b) The real eigenfrequencies of the finite chain as a function of parameter $g_L$ and frequency detuning $\varepsilon$. (c) The enlarged eigenfrequencies of two edge states as a function of parameter $g_L$ without frequency detuning $\varepsilon$. As $g_L$ increases, the splitting edge states gradually coalesce in the EP, which is marked by the black arrow. (d, e) Normalized wave functions of two splitting edge states ($\omega_+$ and $\omega_-$).

## III. EXPERIMENTAL OBSERVATIONS OF THE SENSITIVITY OF TOPOLOGICAL EDGE STATES AT DIFFERENT PHASES AROUND THE EP

To demonstrate experimentally the EP of the edge states, we construct a finite non-Hermitian dimer chain based on the aforementioned theoretical model. The experimental setup is shown in Fig. 4(a). The dimer chain is composed of six resonators. The amplifying, neutral, and lossy resonators are marked, respectively, by G, N, and L in the center of the resonators. This non-Hermitian topological dimer chain is put on a foam substrate with a permittivity near 1. The structure parameters of the resonators are same as those of the composite resonator in Fig. 1. The input port is placed on the left side of the chain to measure the reflection spectrum of the chain. We study the coupling modulation of edge states by tuning the loss in the right lossy resonator while the gain in the amplifying resonator keeps a suitable value $g_G = 0.18$ MHz, as shown in Fig. 3. First we control the gain and loss, balancing $g_G = g_L = 0.18$ MHz, by simultaneously tuning the voltage $U = 1.3$ V and resistance $R = 10$ kΩ for the gain and lossy resonators, respectively. Because of the near-field coupling between two edge states, two splitting edge states will have different eigenfrequencies. The corresponding reflection spectrum is marked A in Fig. 4(b). Then we gradually increase the loss in the lossy resonator and find the point of coalescence of the eigenfrequencies, which corresponds to the EP of the edge states in the non-Hermitian dimer chain. For the EP, the external bias voltage on the amplifying resonator is unchanged at $U = 1.3$ V ($g_G = 0.18$ MHz), whereas the resistor of the lossy resonator is $R = 0.92$ kΩ ($g_L = 0.29$ MHz). The reflectance spectrum corresponding to the EP is marked by B in Fig. 4(b). Next, for greater loss ($g_L > 0.29$ MHz) in the lossy resonator due to further decreasing the value of the resistor, the edge states always degenerate. For example, for the lossy resonator with $R = 0.47$ kΩ ($g_L = 0.43$ MHz), the reflectance spectrum of the topological chain is marked C in Fig. 4(b). Therefore, we observe the phase transition process associated with the EP by tuning the loss of the lossy resonator, which is easily realized by tuning the resistance of the resistor.

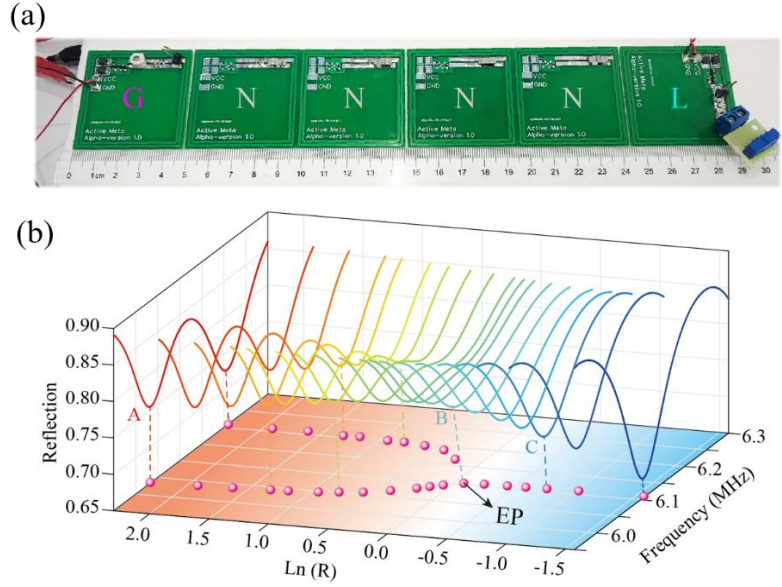

FIG. 4. **Measured reflection spectrum of the 1D non-Hermitian dimer chain.** (a) Photo of the non-Hermitian topological dimer chain. The sample is put on a foam plate 1 cm thick. (b) Measured reflection spectrum as the dissipative loss of the lossy resonator, which is controlled by the tunable resistor at the right end of the chain, increases. Dots denote the frequencies of the edge states. Resistance is given on a logarithmic scale.

To further explore the intriguing properties of the EP in the non-Hermitian topological system, we study the sensitivity of the edge states in three different regimes around the EP. The robustness of edge states in a degenerating regime was recently demonstrated experimentally in a waveguide array with passive-PT symmetry [62]. However, the EP property of edge states in non-Hermitian systems has not been considered. Here we quantitatively study the sensitivity of topological edge states. In particular, the EP for this system is expected to realize a new type of sensor. By changing the loaded lumped capacitor, we add a small frequency perturbation $\varepsilon$ in the lossy resonator. The logarithmic plot of the relation between the frequency shift of the edge states $\delta\omega$ and the capacitance perturbation $\delta C$ is shown in Fig. 5. The EP sensor, marked by B in Fig. 4(b), realized by the edge states exhibits a slope of 1/2 for a small perturbation; the square roots are shown by the green circles in Fig. 5. On one side of the EP, marked by A in Fig. 4(b), the non-Hermitian chain belongs to the splitting region with little loss. The corresponding designed sensor realized by the edge states exhibits a normal slope of 1 for a small perturbation, which is shown by the pink stars in Fig. 5. On the other side of the EP, marked by C in Fig. 4(b), the non-Hermitian chain belongs to the degenerating region. In this case, the edge states are barely affected by the added perturbation, as shown by the blue triangles in Fig. 5. Comparing the results for these three regimes, we see that the sensitivity of edge states to perturbation at the end of the chain is strongly affected around the EP. From the splitting region to the degenerating region, the sensitivity of the edge states first increases and then decreases. It is important to note that although the topological edge states are sensitive to perturbation at the end of chain, they are robust to different perturbations and disorders in the inner chain [11]. Thus, this sensor based on the EP of the edge states in a non-Hermitian topological dimer chain is highly sensitive at the end but topologically protected from internal fabrication errors in the system.

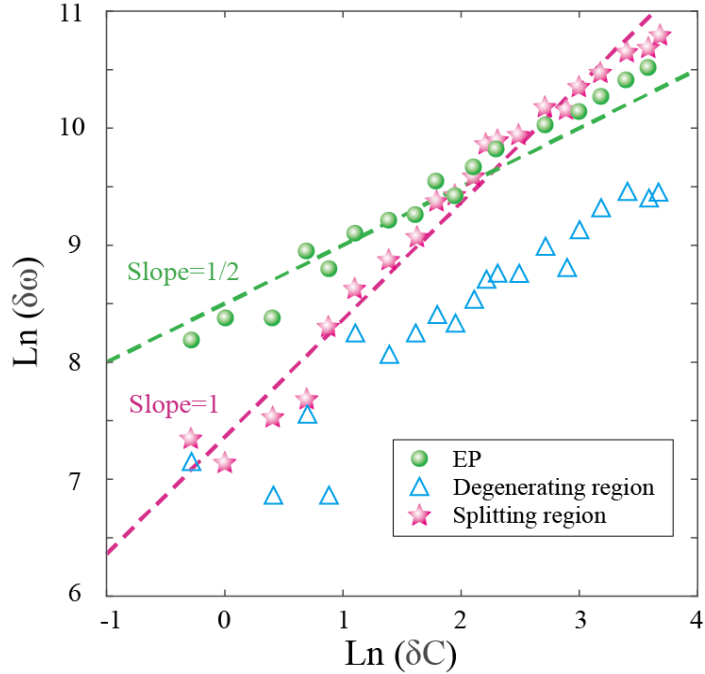

FIG. 5. **Measured frequency splitting of edge states on frequency detuning of the right resonator, which is controlled by the loaded capacitors.** The results are given on a logarithmic scale. The green circles, blue triangles, and pink stars indicate results from the EP, degenerating region, and splitting region, respectively. Green and pink dashed lines with slopes of 1/2 and 1, respectively, are displayed for reference.

## IV. CONCLUSION

In summary, using a finite non-Hermitian topological dimer chain, we study the sensitivity of edge states in three different regimes: the splitting regime, the EP, and the degenerating regime. According to conventional wisdom, the edge states in a topological structure are topologically protected, which makes them robust to structural perturbations. In this work, we show experimentally that the edge states in the degenerating regime after the EP can enhance topological protection in a finite system. However, this scenario breaks down at the EP, and the degenerating regime becomes very sensitive to perturbation at the end of the non-Hermitian chain. Our results for the EP of edge states not only improve understanding of the robustness of topological states but also provide a new scheme for designing a new type of sensor with topological protection against internal disturbances and high sensitivity to boundary perturbations.

## ACKNOWLEDGMENTS

This work was supported by the National Key R&D Program of China (Grant No. 2016YFA0301101), the National Natural Science Foundation of China (NSFC; Grant Nos. 11774261, 11474220, and 61621001), the Natural Science Foundation of Shanghai (Grant Nos. 17ZR1443800 and 18JC1410900), the China Postdoctoral Science Foundation (Grant Nos. 2019TQ0232 and 2019M661605), and the Shanghai Super Postdoctoral Incentive Program.

14. D. A. Dobrykh, A. V. Yulin, A. P. Slobozhanyuk, A. N. Poddubny, and Y. S. Kivshar, Nonlinear control of electromagnetic topological edge states, *Phys. Rev. Lett.* **121**, 163901 (2018).

15. Y. Wang, L. J. Lang, C. H. Lee, B. L. Zhang, and Y. D. Chong, Topologically enhanced harmonic generation in a nonlinear transmission line metamaterial, *Nat. Commun.* **10**, 1102 (2019).

16. S. KruK, A. Poddubny, D. Smirnova, L. Wang, A. Slobozhanyuk, A. Shorokhov, I. Kravchenko, B. Luther-Davies and Y. Kivshar, Nonlinear light generation in topological nanostructures, *Nat. Nanotechnology* **14**, 126 (2019).

17. D. Smirnova, D. Leykam, Y. D. Chong, and Y. Kivshar, Nonlinear topological photonics, *Appl. Phys. Rev.* 7,021306 (2020)..

18. Y. Ota, K. Takata, T. Ozawa, A. Amo, Z. Jia, B. Kante, M. Notomi, Y. Arakawa, and S. Iwamoto, Active topological photonics, arXiv: 1912, 05126 (2019).

19. H. Zhao, P. Miao, M. H. Teimourpour, S. Malzard, R. El-Ganainy, H. Schomerus and L. Feng, Topological hybrid silicon microlasers, *Nat. Commun.* **9**, 981 (2018).

20. M. Parto, S. Wittek, H. Hodaei, G. Harari, M. A. Bandres, J. Ren, M. C. Rechtsman, M. Segev, D. N. Christodoulides, and M. Khajavikhan, Edge-mode lasing in 1D topological active arrays, *Phys. Rev. Lett.* **120**, 113901 (2018).

21. Y. Ota, R. Katsumi, K. Watanabe, S. Iwamoto, and Y. Arakawa, Topological photonic crystal nanocavity laser, *Commun. Phys.* **1**, 86 (2018).

22. R. El-Ganainy, K. G. Makris, M. Khajavikhan, Z. H. Musslimani, S. Rotter and D. N. Christodoulides, Non-Hermitian physics and PT symmetry, *Nat. Phys.* **14**, 11 (2018).

23. L. Feng, R. El-Ganainy, and L. Ge, Non-Hermitian photonics based on parity–time symmetry, *Nat. Photon.* **11**, 752 (2017).